\documentclass[twocolumn]{aastex7}
\usepackage{booktabs}
\usepackage[utf8]{inputenc}
\usepackage{bm}
\usepackage{xcolor}
\usepackage{ulem}
\usepackage{xspace}
\usepackage{float}
\usepackage[T1]{fontenc}
\usepackage{amsmath}
\usepackage{CJK}

\usepackage{ulem}
\definecolor{mygray}{gray}{0.6}
\definecolor{orange}{rgb}{1.0, 0.4, 0.0} 

\newcommand{\fg}[1]{Fig.~\ref{fig:#1}}
\newcommand{\Fg}[1]{Figure~\ref{fig:#1}}

\newcommand{\eq}[1]{Eq.~(\ref{eq:#1})\xspace}

\newcommand{\tb}[1]{Table~\ref{tab:#1}\xspace}
\newcommand{\se}[1]{Sect.~\ref{sec:#1}\xspace}
\newcommand{\ap}[1]{appendix~\ref{ap:#1}\xspace}
\newcommand{\Ap}[1]{Appendix~\ref{ap:#1}\xspace}

\begin{document}
\begin{CJK*}{UTF8}{gbsn}

\title{A Resonant Beginning for the Solar System Terrestrial Planets}

\author[orcid=0000-0002-0054-8880,sname='Shuo Huang']{Shuo Huang (黄硕)}
\affiliation{Department of Astronomy, Tsinghua University, Shuangqing road, Haidian district, 100084, Beijing, China.}
\affiliation{Leiden Observatory, Leiden University, Einsteinweg 55, Leiden, 2333 CC, The Netherlands.}
\email[show]{huangs20@mails.tsinghua.edu.cn}  

\author[orcid=0000-0003-4672-8411,sname='Chris Ormel']{Chris W. Ormel}
\affiliation{Department of Astronomy, Tsinghua University, Shuangqing road, Haidian district, 100084, Beijing, China.}
\email{chrisormel@tsinghua.edu.cn}

\author[orcid=0000-0001-5839-0302,sname='Simon']{Simon Portegies Zwart}
\affiliation{Leiden Observatory, Leiden University, Einsteinweg 55, Leiden, 2333 CC, The Netherlands.}
\email{spz@strw.leidenuniv.nl}

\author[orcid=0000-0002-5486-7828,sname='Eiichiro Kokubo']{Eiichiro Kokubo (小久保 英一郎)}
\affiliation{National Astronomical Observatory of Japan, Osawa, Mitaka, 2-21-1, Japan.}
\email{kokubo.eiichiro@nao.ac.jp}

\author[orcid=0009-0006-2919-2394,sname='Tian Yi']{Tian Yi (易天)}
\affiliation{Department of Astronomy, Tsinghua University, Shuangqing road, Haidian district, 100084, Beijing, China.}
\email{yit23@mails.tsinghua.edu.cn}

\begin{abstract}
    In the past two decades, transit surveys have revealed a class of planets with thick atmospheres -- sub-Neptunes -- that must have completed their accretion in protoplanet disks. When planets form in the gaseous disk, the gravitational interaction with the disk gas drives their migration and results in the trapping of neighboring planets in mean motion resonances, though these resonances can later be broken when the damping effects of disk gas or planetesimals wane. It is widely accepted that the outer Solar System gas giant planets originally formed in a resonant chain, which was later disrupted by dynamical instabilities. Here, we explore whether the early formation of the terrestrial planets in a resonance chain (including Theia) can evolve to the present configuration. Using N-body simulations, we demonstrate that the giant planet instability would also have destabilized the terrestrial resonance chain, triggering moon-forming giant impacts in 20--50\% of our simulated systems, dependent on the initial resonance architecture. After the instability, the eccentricity and inclination of the simulated planets match their present-day values. Under the proposed scenario, the current period ratio of 3.05 between Mars and Venus -- devoid of any special significance in traditional late formation models -- naturally arises as a relic of the former resonance chain. 

\end{abstract}


\section{Introduction and motivation} 
If two planets are in a (first-order) orbital mean motion resonance, their period ratio stays close to an integer ratio, $j+1{:}j$. Because the conjunction point occurs close to the pericenter of the inner and apocenter of the outer planet, the mean motion resonance enhances the dynamical stability of planetary systems \citep{TamayoEtal2017, GoldbergEtal2022}. {Mathematically, the resonance angles, which express the point where in the orbit conjunctions take place, are said to} librate for planets in resonance. Examples in our solar system are the 1:2:4 resonance among Galilean moons and the 3:2 resonance between Neptune and Pluto. Many exoplanet systems are discovered to be in resonance as well, e.g., TRAPPIST-1 \citep{GillonEtal2017, LugerEtal2017}, TOI-178 \citep{LeleuEtal2021}, TOI-1136 \citep{DaiEtal2022}, HD110067 \citep{LuqueEtal2023}. Planets in resonances are expected to experience large-scale migration, which is naturally explained by the formation of planets early, in a gas-rich disk \citep{PapaloizouSzuszkiewicz2005}. Exoplanet formation is widely believed to occur within protoplanetary disks, as supported by both observations \citep{AndrewsEtal2018, KepplerEtal2019, BarberEtal2024} and theoretical models \citep{DrazkowskaEtal2023, MordasiniEtal2015}. Efficient Type I migration often traps planets into mean motion resonances, as evidenced by the observed period ratio distributions in exoplanet populations \citep{HuangOrmel2023i, HamerSchlaufman2024}, particularly in young systems \citep{DaiEtal2024}. {In contrast, non-resonant planets with circulating resonance angles located near integer period ratios are much less stable \citep{HuEtal2025}.} 

\begin{figure*}
	\centering
	\includegraphics[width=0.8\linewidth]{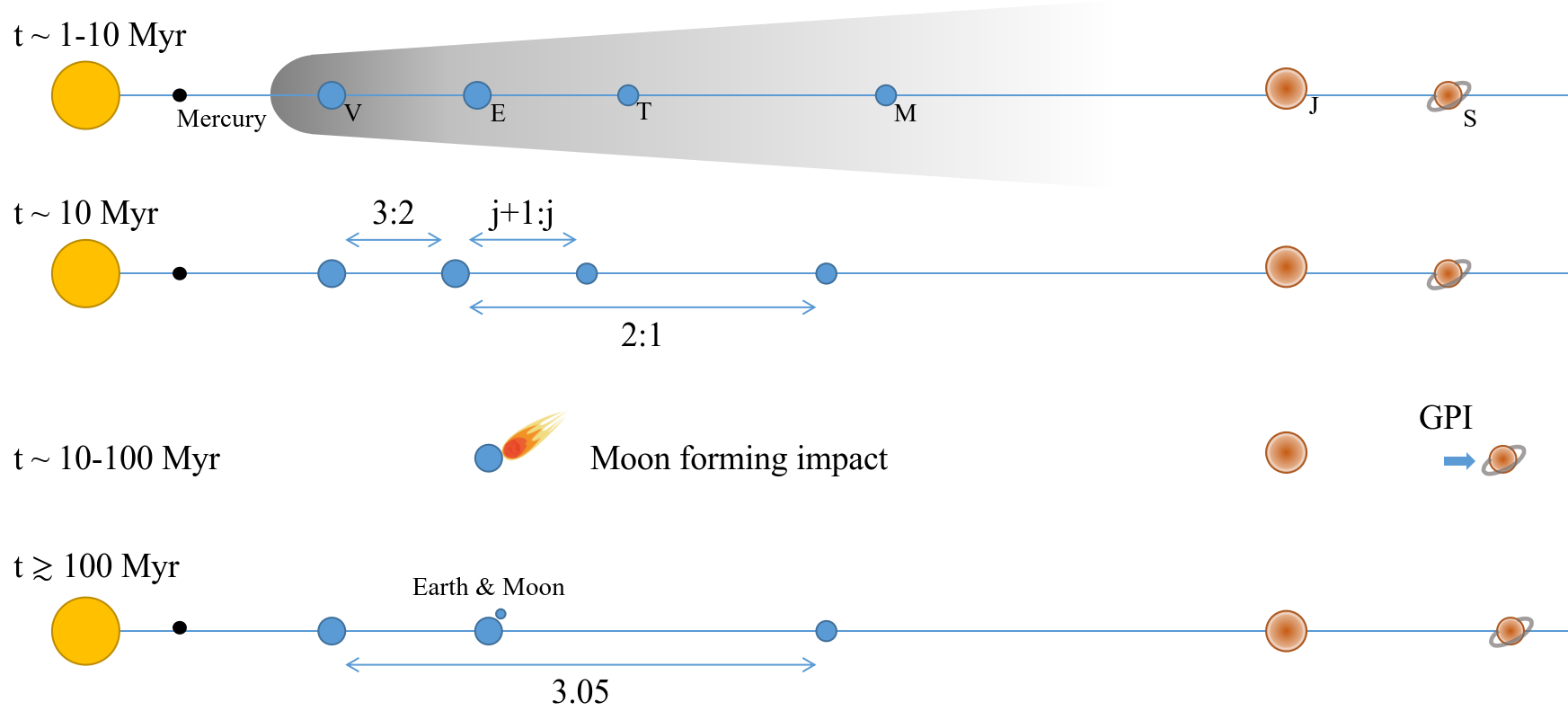}
    \caption{{Sketch illustrating the dynamical evolution of the solar system planets during and after disk dissipation.} The two-sided arrows represent the existence of mean motion resonances. Blue and brown scatters represent terrestrial bodies, including Venus (V), proto-Earth (E), Theia (T), and Mars (M), and gas giants including Jupiter (J) and Saturn (S). In the disk phase ($t\lesssim$1 Myr), terrestrial planets form and migrate convergently. Theia and Earth are in $j+1:j$ resonance, where $j$ is the resonance number. At $t\gtrsim1$ Myr, i.e., just after disk dispersal, planets are still locked in a resonance chain. At $t\sim10$ Myr, the outward migrating Saturn excites Jupiter, triggering the giant planet instability (GPI), which also destabilizes the resonance chain and triggers the moon-forming giant impact. After $t\sim100$ Myr, the planets stabilize to the current architecture. }
	\label{fig:sketch}
\end{figure*}

Giant planets in our solar system are widely believed to start in a resonant configuration \citep{TsiganisEtal2005}. The resonance chain was then broken during giant planet instability \citep[GPI,][]{LiuEtal2022, GriveaudEtal2024, BrownEtal2024}. 
GPI is widely recognized as a turning point in the early evolution of our Solar System. In particular, such an event plays a key role in reproducing the observed orbital properties of the outer solar system bodies, such as the eccentricities and inclinations of the giant planets \citep{TsiganisEtal2005}, distribution of the Kuiper Belt \citep{LevisonEtal2008} and asteroid belt \citep{O'BrienEtal2007}, Trojan asteroids of Jupiter and Neptune \citep{MorbidelliEtal2005}, and irregular moons around giant planets \citep{NesvornyEtal2014}. However, the migration of giant planets during GPI not only affects their own orbits but also has profound consequences for any existing terrestrial planets. As Saturn migrates outward, it induces sweeping secular resonances that can perturb the orbits of inner planets, potentially triggering orbital crossing and collisions \citep{BrasserEtal2009, KaibChambers2016, ThommesEtal2008}.

Dynamical constraints indicate that the GPI occurred within 100 Myr after the formation of the Solar System \citep{NesvornyEtal2018, deSousaEtal2020}. This timing suggests that terrestrial planet formation was largely complete before the instability took place \citep{KobayashiDauphas2013}. Consequently, the terrestrial planets would have been in place when GPI perturbed their orbits, setting the stage for significant dynamical rearrangement and collisions. The most compelling evidence of such a collision in our inner Solar System must be the Moon-forming giant impact \citep{CanupAsphaug2001}.
Cosmochemical studies constrain the Moon-forming event to have occurred between 40 and 120 Myr after Solar System formation \citep{BarboniEtal2017, ThiemensEtal2019}, which overlaps significantly with the estimated timing of the GPI. This temporal correlation has motivated studies suggesting that the dynamical disturbances associated with the GPI may have triggered the Moon-forming giant impact \citep{ClementEtal2023}. Specifically, in their model terrestrial planets form after gas dispersal by collisional growth between Mars-sized embryos \citep{ClementEtal2019i}.  

It is still under debate whether terrestrial planets largely complete their growth early in the protosolar nebulae, when gas was still present \citep{JohansenEtal2024, MorbidelliEtal2024i}. Here we hypothesize that the terrestrial planets in our Solar System formed early in the protosolar nebulae---a scenario supported by both formation models \citep{JohansenEtal2021, BrovzEtal2021} and observational work {on Mars isotopes \citep{KobayashiDauphas2013} and Earth oxidation state \citep{YoungEtal2023, JohansenEtal2023}}. Naturally, planets can get trapped in resonance due to disk migration. However, it remains unclear whether the observed architecture of the terrestrial planets can be reproduced starting from an initial resonance chain. To investigate this hypothesis, this work initializes terrestrial planets in a resonance chain. Their long-term dynamical evolution is tracked and studied. 

The paper is structured as follows. We detailed the methodology in \se{method}. Our main results and their comparison to the present solar system are in \se{main}. The implications of our findings are discussed in \se{implication}. Finally, we state our conclusions in \se{conclusion}.


\begin{table*}
\label{tab:solar}
\caption{Orbital properties of Solar system planets\citep{MurrayDermott1999}. The planet mass, semimajor axis, orbital period, current eccentricity, and inclination in the elliptical plane and invariable plane are listed. }\label{tab1}%
	\begin{tabular}{@{}lllllll@{}}
		\toprule
		Name    & $m_\mathrm{p}$ [$M_\oplus$] & $a_\mathrm{p}$ [au] & $P$ [days] & $e$ &  Ecliptic $i$[deg] & Invariable $i$ [deg]\\
		\midrule
		Venus   & 0.82              & 0.72     & 224.7         & 0.0068       & 3.394        & 2.19     \\
		Earth   & 1                 & 1        & 365.2         & 0.0167       & 0.000        & 1.57     \\
		Mars    & 0.11              & 1.52     & 687           & 0.0934       & 1.850        & 1.67     \\

		Jupiter & 317.8             & 5.20     & 4331          & 0.0484       & 1.303        & 0.32     \\
		Saturn  & 95.2              & 9.57     & 10,747        & 0.0541       & 2.485        & 0.93     \\
		\bottomrule
	\end{tabular}
\end{table*}

\section{Model}
\label{sec:method}
We illustrate the dynamical evolution of terrestrial planets in \fg{sketch} over time. In our setup, we include four terrestrial planets—Venus, proto-Earth, Theia, and Mars—in resonance. {We exclude Mercury from the resonance chain because of its potentially distinct formation history compared to the other terrestrial planets \citep[e.g.][]{JohansenDorn2022, MorbidelliEtal2022}. Mercury is also very far from other terrestrial planets, interior to the 3:1 period ratio of Venus. Including it would not affect the current results due to the very low mass of Mercury (half that of Mars). Besides, Mercury seems to experience more complex dynamical interaction with the Sun, as evidenced by the 3:2 spin-orbit resonance \citep[e.g.][]{WieczorekEtal2012, NoyellesEtal2014}. } 

In addition, we include two gas giants, Jupiter and Saturn, initially arranged more compactly than observed today with a period ratio smaller than 2, and Saturn migrates outward to its present orbit \citep{MorbidelliEtal2007, LevisonEtal2011}. During this migration phase (giant planet instability), Jupiter's eccentricity is excited at various resonance locations, and the terrestrial planets experience secular perturbations from the giant planets \citep{BrasserEtal2009, KaibChambers2016}. 

In our scenario, the terrestrial resonance chain is broken during Saturn's outward migration, leading to increased eccentricities among the terrestrial planets. The Moon-forming giant impact is then triggered once the orbits of proto-Earth and Theia overlap. Since Venus and Mars are relatively distant from their neighboring terrestrial planets, their orbits remain largely unchanged, thereby preserving the initial 3:1 period ratio between Mars and Venus, which is consistent with current observations. 

We use the REBOUND \citep{ReinLiu2012} package to conduct the N-body calculations. The migration and eccentricity damping forces are implemented using REBOUNDx \citep{TamayoEtal2020} and the MERCURIUS integrator \citep{Chambers1999, ReinEtal2019} is employed. 

\subsection{Initial conditions}
The terrestrial planets are initially set in resonance in our N-body simulations. {In our model, Venus reaches the migration barrier, possibly the disk inner edge, which promotes convergent migration and thus resonance trapping (see \fg{sketch}).} To achieve these initial resonant conditions, we follow the approach of \citep{TamayoEtal2017}: we first position the planets slightly away from exact resonance and then allow convergent migration to capture them into resonance while damping their eccentricities. The acceleration term due to the gravitational interaction with the disk used in our N-body calculations is
\begin{equation}
    \boldsymbol{F_i} = \frac{\boldsymbol{v_i}}{2\tau_{\mathrm{a},i}}+\frac{2(\boldsymbol{v_i} \cdot \boldsymbol{r_i})\boldsymbol{r_i}}{r_i^2\tau_{\mathrm{e},i}},
\end{equation}
where $\boldsymbol{v_i}$ and $\boldsymbol{r_i}$ denote the velocity and position vector of planet $i$, and $\tau_{\mathrm{a},i}$ and $\tau_{\mathrm{e},i}$ are the semimajor axis and eccentricity damping timescale. 
Venus migrates outward on a timescale $\tau_\mathrm{a}$ of 10 Myr. Such a long timescale represents the end stage of the disk environment. Planet eccentricities are damped on a timescale of $\tau_\mathrm{e}=\tau_\mathrm{a}/K_\mathrm{e}$, where $K_\mathrm{e}$ is a free parameter ranging from $10^2$ to $10^4$. The parameter $K_\mathrm{e}$ controls the eccentricities of the terrestrial planets in the resonance chain \citep{TanakaWard2004}. Large $K_\mathrm{e}$ leads to smaller eccentricities. 

{The simple migration setup described above ensures convergent migration and, consequently, resonance trapping in a controlled manner. In reality, however, the process of resonance trapping is more complex. Convergent migration may occur, for instance, if Venus halts its migration at the migration barrier of the protoplanet disk \citep{LiuEtal2022, OgiharaEtal2024i, WuChen2025}. The detailed processes by which terrestrial planets form and migrate within the disk remain uncertain but do not matter for this study as long as the terrestrial planets end up in the resonance chain depicted in \fg{sketch}}. 

Jupiter and Saturn are also included in the simulation. Jupiter is initialized at its present-day orbital period with eccentricity $e_\mathrm{J,\,ini}$ ranging from 0 to 0.05, and an inclination of 0.5 degrees.  Saturn is placed at an initial period ratio of 1.9 with Jupiter, i.e., closer to Jupiter than at present and within the 2:1 resonance location.

Most simulations succeed in forming a resonance chain, characterized {by the libration of resonance angles among the terrestrial planets}. We consider the point when Venus has migrated to 0.72 au (the current location of Venus) as the initial conditions for the post-disk dynamical evolution. 

\subsection{Giant planet instability}
During the phase of giant planet instability, Saturn migrates outward. The driven process and the timescale of such migration are still under active debate \citep{LiuEtal2022, deSousaEtal2020, AgnorLin2012}. We therefore parameterize Saturn's migration with the additional acceleration term in the N-body calculation:
\begin{equation}
    \boldsymbol{F}=\frac{1}{2}\frac{a_0-a_f}{a}\frac{\boldsymbol{v}}{\tau_\mathrm{S}}e^{-t/\tau_\mathrm{S}},
\end{equation}
where $a_0$ and $a_f$ are the initial and expected final positions of the body, $a$ and $\boldsymbol{v}$ are the semimajor axis and velocity of the migrating body, and $\tau_\mathrm{S}$ controls the timescale of the migration. It results in the semimajor axis changing in the form of $a(t)=a_f+(a_0-a_f)e^{-t/\tau_\mathrm{S}}$ \citep{BrasserEtal2009, KaibChambers2016, FangEtal2025}.
Such exponential migration has been implemented in REBOUNDx by \citet{Ali-DibEtal2021}. We run simulations varying the outward migration timescale of Saturn $\tau_\mathrm{S}$ from 0.8 to 16 Myr. The solar system is integrated for 100 Myr. The parameter sensitivity is discussed in \Ap{parameter}

\begin{figure*}
	\centering
    \includegraphics[width=1.6\columnwidth]{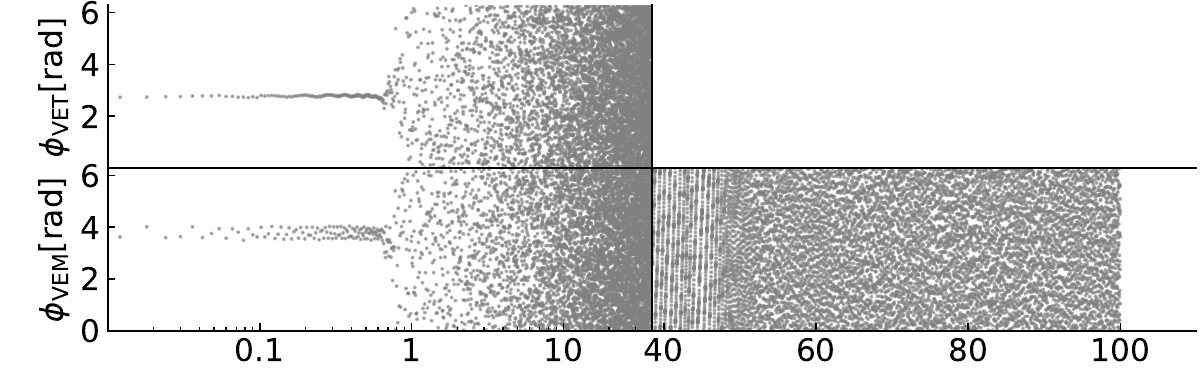}
	\includegraphics[width=1.6\columnwidth]{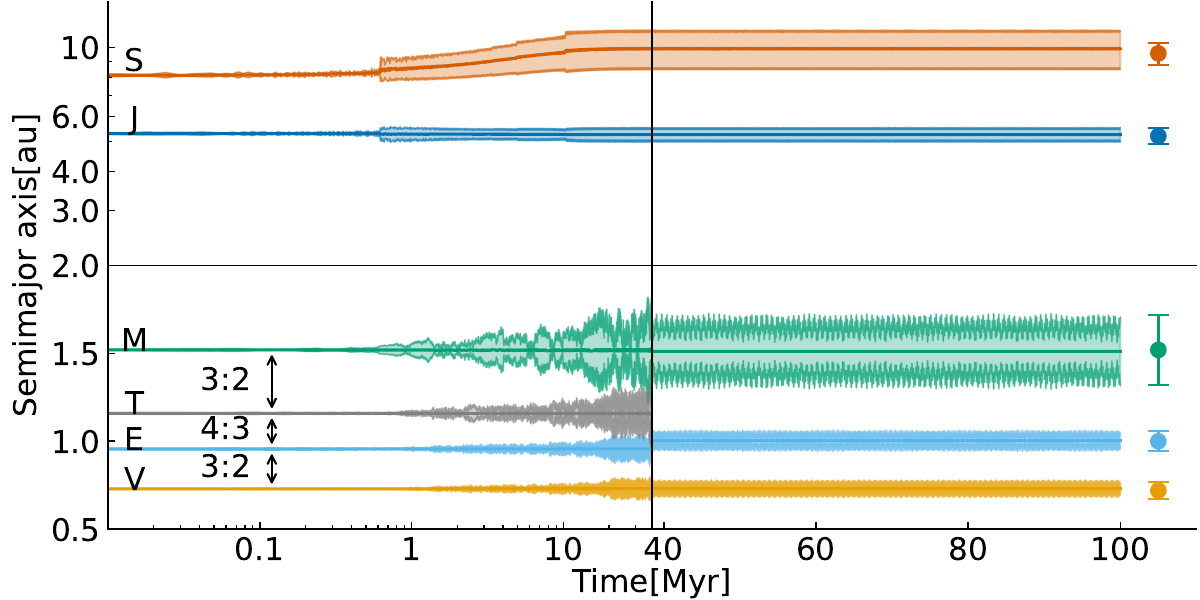}
    \caption{Dynamical evolution of planets during Saturn's outward migration and the post-Moon-forming impact phase. {The top panel shows the evolution of 3-body resonance angles of Venus, Earth, and Theia ($\phi_\mathrm{VET}$) and Venus, Earth, and Mars ($\phi_\mathrm{VEM}$). The resonance angles are defined in \se{main}. The bottom panel shows the changes of planet semimajor axis and eccentricity with time.} The curves show the orbital evolution of each body, including its semi-major axis (thick), perihelion, and aphelion (thin). The terrestrial planets are initially in a resonance chain. The corresponding resonances are labeled by the double-headed arrows. The semimajor axis, aphelion, and perihelion of the current solar system planets are denoted by the error bars at the right. The adopted parameters are: $e_\mathrm{J, ini}=0.005,\, \tau_\mathrm{S}=4.7\,\mathrm{Myr}$ and $K_\mathrm{e}=5.6\times10^3$. The crossing of the 2:1 resonance between Saturn and Jupiter (${\approx}1\,\mathrm{Myr}$) also destabilizes the inner solar system, breaking its resonances and resulting in the Moon-forming event (${\approx}35\,\mathrm{Myr}$).}
	\label{fig:example}
\end{figure*}

\subsection{Theia and proto-Earth mass-radius calculation}
After Theia collides with proto‐Earth, the resulting merger—the Earth-Moon system—is located at the center of mass of the two original bodies. Since Earth currently orbits at 1 au, we fix the center of mass at 1 au in our setup, meaning that different initial configurations for Earth and Theia yield different planet mass ratios.

Assuming a perfect merger and small eccentricities, angular momentum conservation gives:
\begin{equation}
	m_\mathrm{E}\sqrt{a_\mathrm{E}}+m_\mathrm{T}\sqrt{a_\mathrm{T}}=(m_\mathrm{E}+m_\mathrm{T})\sqrt{a_\oplus}.
\end{equation}
The "E" and "T" notations are for proto-Earth and proto-Theia, while $\oplus$ notation is for the present-day Earth. After inserting values, we find
\begin{equation}
\label{eq:massratio}
m_\mathrm{T} =m_\mathrm{E}\frac{\sqrt[3]{\frac{2P_\oplus}{3P_{\mathrm{V}}}}-1}{\sqrt[3]{\frac{j+1}{j}} -\sqrt[3]{\frac{2P_\oplus}{3P_{\mathrm{V}}}}}  
    =m_\mathrm{E}\frac{0.0272}{\sqrt[3]{\frac{j+1}{j}}-1.0272}
\end{equation}
where $j$ is the resonance number of proto-Earth and Theia. When $j=4$, $m_\mathrm{T}=0.54m_\mathrm{E}$. when $j=3$, $m_\mathrm{T}=0.37m_\mathrm{E}$. We assume $m_\mathrm{E}+m_\mathrm{T}=1.05m_\mathrm{\oplus}$ in all of our calculations. Venus and Mars do not experience giant impacts as suggested for Earth, and their mass does not change in our model. 

Planets collide if their physical radii overlap. Planet radii are adopted from current measurements, except for proto‐Earth and Theia, which no longer exist in the present Solar System. For bodies with masses below $1\,m_\oplus$, we use the mass-radius relationship 
\begin{equation}
	\frac{R}{1\,R_{\oplus}}=\left(\frac{m}{1\,m_\oplus}\right)^{0.29}
\end{equation}
as given in \citep{SeagerEtal2007}, which employs a polytropic model to describe the internal structure of rocky planets. The adopted planet properties are summarized in \tb{solar}.

\section{Comparison with Solar system}
\label{sec:main}
Firstly, we hypothesize that the four terrestrial planets—Venus, Earth, Theia, and Mars—emerged from the gaseous disk in a 2:3:4:6 resonance chain. We find that this resonant architecture is a natural consequence regardless of whether planet formation proceeds via pebble accretion or planetesimal accretion (see \Ap{formation}). This resonance chain guarantees a 3:1 period ratio between Mars and Venus. Due to their close spacing, Theia and proto-Earth are prone to collide once the resonance is broken, giving rise to the Earth-Moon system at 1 au. Consequently, the mass ratio between Theia and proto-Earth is based on their initial positions; with Venus fixed at its current distance of 0.72 au, the mass ratio amounts to 0.37 (see \eq{massratio}). 

\begin{figure*}
	\centering
	\includegraphics[width=0.8\linewidth]{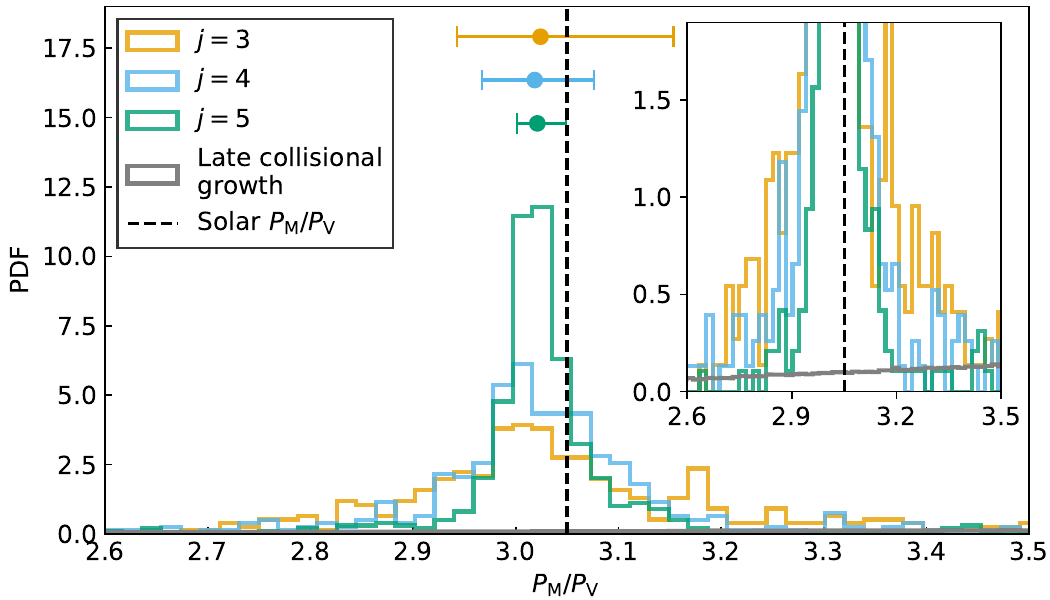}
    \caption{{Mars-Venus period ratio probability distribution from different models}. The current value is shown by the vertical dashed line. The colored histograms are the subgroup of simulated systems (defined in \fg{fin_ej}) of the early, resonant, formation models, with different colors representing the Theia and Earth resonances through the resonance number $j$. The corresponding median 50\% ranges and the median values are indicated by the error bars and their center values. The grey histograms are the results of the post-nebulae collisional growth model, which does not feature an elevated probability near the 3:1 period ratio (see \Ap{latecollision}). The figure inset features a shorter range of the y-axis}. 
	\label{fig:period_ratio_comp}
\end{figure*}

It was constrained that the giant planet instability occurred in the first $10-100$ Myr after the formation of Calcium-aluminum-rich Inclusions (CAIs) \citep{EdwardsEtal2024}. This instability may be triggered either by gas disk dissipation \citep{LiuEtal2022, ThommesEtal2008}, by self-driven dynamical processes \citep{deSousaEtal2020, GriveaudEtal2024}, or even by stellar encounters \citep{PortegiesZwartEtal2021, BrownEtal2024}. In our simulations, we simply migrate Saturn outward, following approaches used in previous studies \citep{BrasserEtal2009, KaibChambers2016, FangEtal2025}. The timescale of Saturn's outward migration, $\tau_\mathrm{S}$, is parameterized between 0.8 and 16 Myr \citep{LiuEtal2022, GriveaudEtal2024}. Although the dynamical instability of the giant planets has shaped many other features of the outer Solar System \citep[e.g.][]{TsiganisEtal2005, MorbidelliEtal2005, OrmelHuang2025}, here we focus on its effect on the architecture of the terrestrial planets.

\Fg{example} illustrates a representative simulation in which the primordial resonance chain is broken by the migrating Saturn. {Initially, both 3-body resonance angles, which are composed of the mean longitudes of relevant planets $\phi_\mathrm{VET}$ and $\phi_\mathrm{VEM}$ librate, where $\phi_\mathrm{VET}=2\lambda_\mathrm{V}-6\lambda_\mathrm{E}+4\lambda_\mathrm{V}$ and $\phi_\mathrm{VEM}=2\lambda_\mathrm{V}-4\lambda_\mathrm{E}+2\lambda_\mathrm{M}$.} In our model, Jupiter's eccentricity is excited to approximately $e_\mathrm{J}\approx0.05$ as Saturn crosses the 2:1 resonance \citep{LithwickEtal2012}. Subsequently, the $g_5$ secular resonance \citep{BrasserEtal2009} sweeps through the terrestrial planets' orbits and the angular momentum deficit (AMD) -- a measure of the dynamical excitation of the system \citep{Laskar1997} -- diffuses \citep{KaibChambers2016}, disrupting their resonant configuration. {With circulating resonance angles planets are no longer in resonance, but remain near integer period ratios.} {In addition to the secular perturbations from Jupiter, close encounters among near-resonant terrestrial planets further amplify planet eccentricities}, leading to orbital overlap between Theia and Earth, culminating in their collision. Following the Moon-forming event, the terrestrial orbits stabilize as they are displaced away from both secular and mean-motion resonances. Although the semi-major axes remain essentially constant, the eccentricities continue to librate under Jupiter's secular influence \citep{MurrayDermott1999}. Overall, our simulation reproduces the orbital properties of the present-day terrestrial planets.

\begin{figure*}
	\centering
	\includegraphics[width=0.8\linewidth]{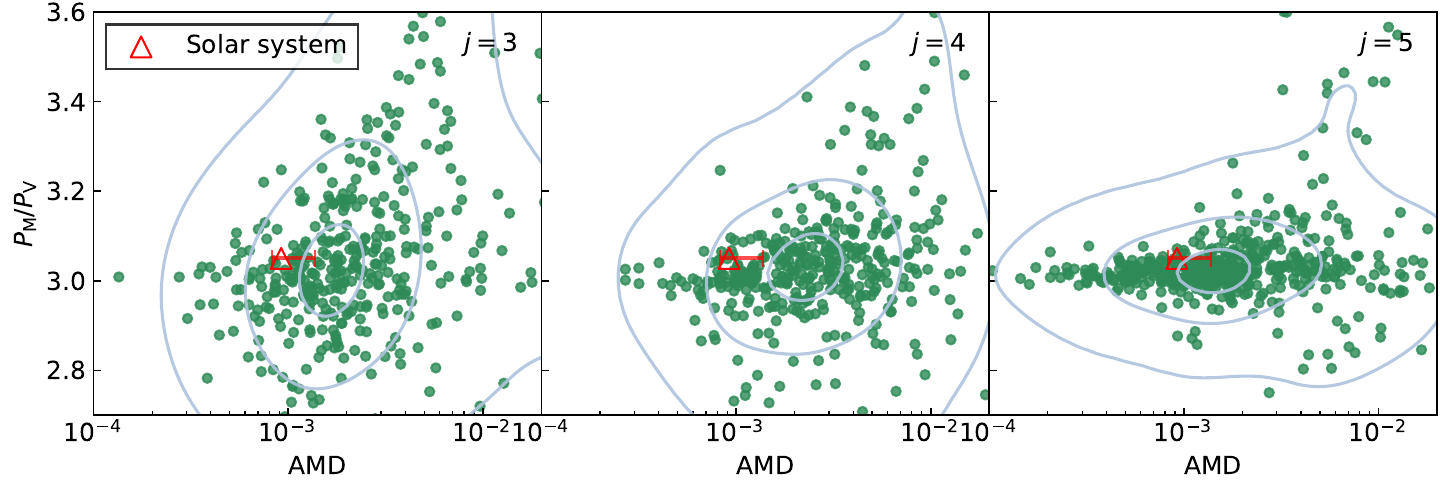}
    \includegraphics[width=0.8\linewidth]{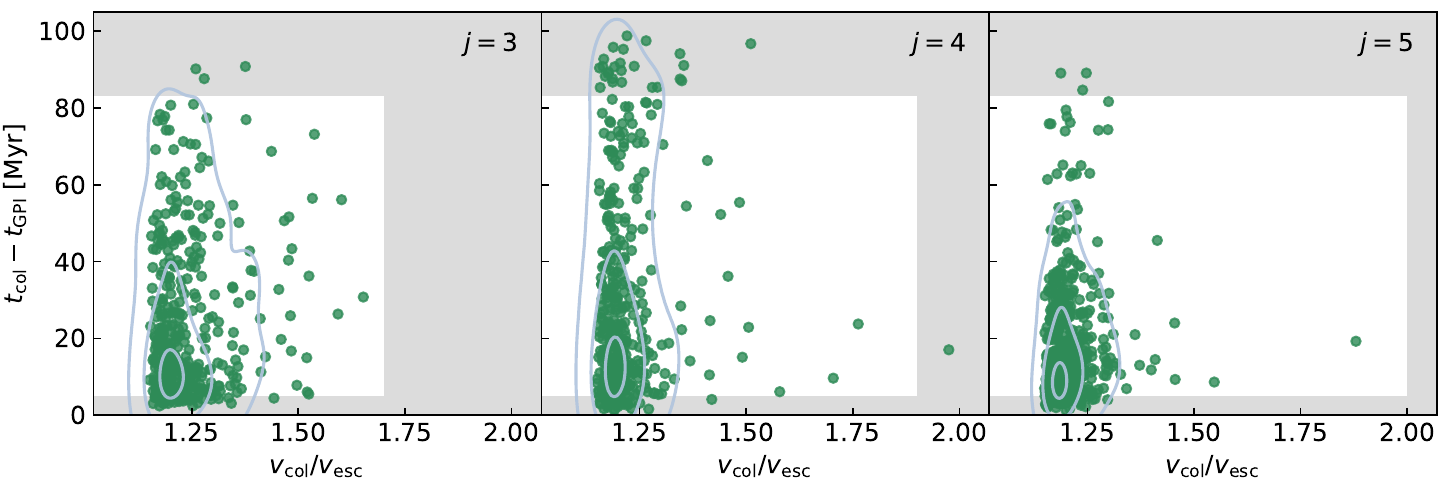}
    \caption{Planet properties for the simulated Solar-like systems in which Theia collides with Earth. The definition of Solar-like system is in\fg{fin_ej}. Each simulation starts with a terrestrial resonance chain and two gas giants. Theia and Earth are initially in $j+1{:}j$ resonance, with the resonance number $j$=3, 4, or 5 labeled on the top right of each panel. The top panels show the Angular Momentum Deficit (x-axis) and Mars-Venus period ratio (y-axis). The present-day Solar System is marked by a triangle for comparison, while the error bar shows the range in secular evolution \citep{ItoTanikawa2002}. The bottom panels show Theia's collision velocity with Earth in terms of the escape velocity (x-axis) and the collision time with respect to the GPI (y-axis). Permissible cosmochemical constraints on the timing \citep{BarboniEtal2017, ThiemensEtal2019} of the GPI and SPH simulation constraints on the impact velocity of the Moon-forming event \citep{TimpeEtal2023} are indicated by the white window, in which the GPI is assumed to occur at 37 Myr after CAI \citep{deSousaEtal2020}.}
	\label{fig:AMD_PMPV}
\end{figure*}

Perhaps the most direct quantity to compare the simulated systems with the current Solar System is the period ratio between Venus and Mars, which is observed to be $P_\mathrm{M}/P_\mathrm{V}=3.057$. We compile the period ratios from our simulations across different parameter sets. The statistics of simulated systems, such as displayed in \fg{period_ratio_comp}, include only those simulations in which Jupiter's free and forced eccentricities match their present-day values. See \ap{parameter}(\fg{fin_ej}) for discussion on how this selection was made. All other figures in the main text are also based on this conditional subset. The probability distribution of $P_\mathrm{M}/P_\mathrm{V}$ peaks at approximately 3.01, consistent with the planets' pre-instability configuration---a value that arises from the balance between planet migration and eccentricity damping \citep{CharalambousEtal2022}. 
Due to planet-planet scattering during the simulation, the distribution broadens from a $\delta$-function, with the observed period ratio falling within the 25\% range from the median. For comparison, we also overplot the period ratio distribution from the late formation model (see \fg{period_ratio_comp_zoom} in \Ap{latecollision}), in which terrestrial planets grow via mutual giant impacts between planet embryos. Compared to the late formation model, the early formation scenario yields a probability density higher by more than an order of magnitude of reproducing the observed Mars-Venus period ratio. The zoomed-in version of the period ratio distribution in the late formation scenario is also shown, in the subset.  

We also run two additional sets of simulations, varying the resonant configurations of Theia and proto‐Earth by initializing them in a 4:5 mean‐motion resonance (MMR, $j$=4) and a 5:6 MMR ($j$=5), while preserving the 2:3:6 resonance chain among Venus, proto‐Earth, and Mars to maintain the 3:1 period ratio between Mars and Venus. In each case, we recalculate the mass ratio of Theia-to-proto‐Earth to ensure that their center of mass remains at 1\,au. As the resonance index $j$ increases, Theia’s mass also increases. The resulting Mars-Venus period ratios are shown in \fg{period_ratio_comp} (blue and green lines). When Theia starts in a closer orbit to Earth (i.e., with a higher $j$ value), the spread of the period ratio distribution narrows. This occurs because a more massive Theia is less susceptible to excitation and remains closer to Earth, making it more difficult to perturb Mars. Consequently, the simulated Mars-Venus period ratio exhibits a stronger peak around 3.01, although the observed value still falls within the quartile range of the distribution. Later dynamical processes could further adjust the planet positions. For example, the $\approx$0.05 wt\% late veneer accreted by Mars \citep{BrasserEtal2016} may have slightly shifted its orbit outward.

Apart from the period ratio, we used another system-level indicator to test whether our simulated systems are consistent with the global properties of the Solar System -- the Angular Momentum Deficit (AMD)
\begin{equation}
	\mathrm{AMD} = \frac{\Sigma_jM_j\sqrt{a_j}(1-\sqrt{1-e_j^2}\cos{i_j})}{\Sigma_jM_j\sqrt{a_j}}. 
\end{equation} 
Our simulation results match the Solar System well. As shown in \fg{AMD_PMPV} (top panels), the solar system resides within the central 10\% of the distribution derived from our simulations. The higher eccentricity of Jupiter (grey scatter) results in systems with larger terrestrial AMD. A more direct approach is to individually compare each planet’s eccentricity and inclination. As demonstrated in \fg{inc_ecc}, the terrestrial planets in our simulations exhibit eccentricities and inclinations similar to those observed in the present-day Solar System.

\begin{figure*}
	\centering
	\includegraphics[width=0.7\linewidth]{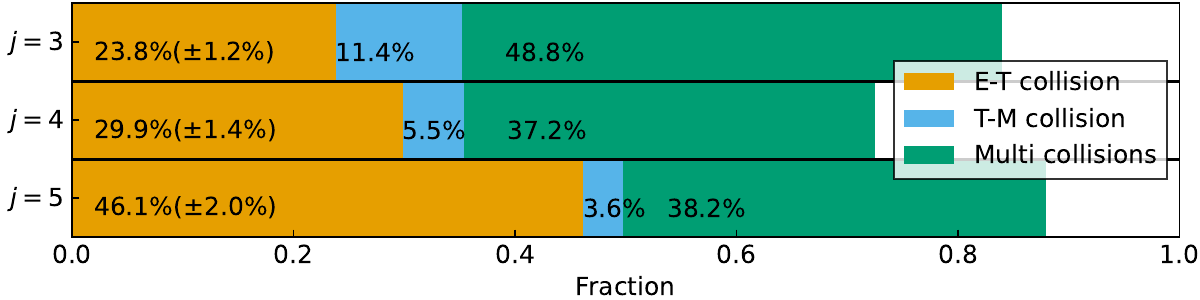}
    \caption{{Statistical outcomes of simulated Solar-like systems, with final Jupiter eccentricities similar to the present day value (\fg{fin_ej}).} The architecture of simulated systems is classified into four groups: Theia collides with Earth (orange); Theia collides with Mars (blue); There are multiple collisions (green); No collisions occur (white). The fractions of different types of simulated Solar systems are stated, where for the Earth-Theia collision case the Poisson error is also stated. Different $j$ values (y-axis) represent different initial resonance configurations between Earth and Theia. 
    }
	\label{fig:parameter_study}
\end{figure*}

The simulated planet systems exhibit a wide range of dynamical outcomes. In our simulations, Theia does not always collide with Earth; it may instead impact Mars, or if the system becomes too dynamically excited, it may lead to multiple planet collisions. Conversely, in some simulations, the planets remain on their initial orbits with minimal excitation, or they become over-excited to high inclinations that no collisions occur at all. \Fg{parameter_study} presents the statistical distribution of these outcomes. The desired scenario—Theia colliding with Earth—occurs in 20-50\% of simulated systems, depending on the initial resonant configuration ($j$). Naturally, placing Theia in a mean-motion resonance (MMR) closer to Earth promotes orbital overlap between the two bodies, while simultaneously reducing the probability of Theia impacting Mars.

It is commonly accepted that the Moon was formed through a giant impact \citep{CanupEtal2023i}. We record the time of the Theia-Earth impact in our simulations in \fg{AMD_PMPV}. About 50\% of our simulated systems have Theia and Earth colliding within $30$ Myr. Since our simulations start with the onset of giant planet instability (GPI), the absolute time (w.r.t CAI formation) of the Theia-Earth collision needs to be corrected accordingly. Dynamical evidence of Kuiper belt objects in our solar system shows that the GPI occurs at 37-62 Myr after CAI \citep{deSousaEtal2020}. Assuming GPI occurs at 37 Myr, then the simulated Theia and Earth collide at $\approx67$ Myr w.r.t CAI formation. Such timing is consistent with the cosmochemical constraints that the Moon-forming giant impact is at 40-120 Myr \citep{BarboniEtal2017, ThiemensEtal2019}. 

In the canonical model of the Moon-forming event, a Mars-sized body collides with proto-Earth, delivering the angular momentum required for the Earth-Moon system \citep{CanupAsphaug2001}. However, the isotopic similarities between the Earth and the Moon \citep{WiechertEtal2001} suggest that either the impactor had nearly the same composition as Earth or, more naturally, that the collision was highly energetic. For this reason, a scenario with a more massive impactor has recently emerged as a promising alternative \citep{LockEtal2018, AsphaugEtal2021}, which also guarantees a sufficiently massive proto-lunar disk \citep{TimpeEtal2023}. In our model, the mass ratio $\gamma$ between Theia and proto-Earth is 0.37, 0.54, and 0.76 for the initial 4:3, 5:4, and 6:5 Theia-Earth resonances, respectively. Given these mass ratios, the collision velocity is constrained to $\lesssim 1.5 v_\mathrm{esc}$, where $v_\mathrm{esc}$ is the mutual escape velocity of the target and impactor \citep{TimpeEtal2023, AsphaugEtal2021}. As shown in \fg{AMD_PMPV}, our simulated systems exhibit relatively low impact velocities because the resonance configuration allows the planets to be closely spaced initially. These modest impact velocities are in line with the aforementioned constraints in the scenario of a more massive Theia than the canonical model. 

\section{Implications}
\label{sec:implication}
The implication of our model is that proto-Earth formed early, {which has been argued to be inconsistent with the observed Hf-W isotope system in bulk silicate Earth (BSE) \citep{MorbidelliEtal2024i}}. This occurs because, during core formation, the {lithophile} parent element $^{182}$Hf remains in the mantle, leading to an excess of its decay product $^{182}$W relative to present-day measurements \citep{OlsonEtal2022}.
However, several factors could operate to decrease the relative abundance of
$^{182}$W in the Earth's mantle. For example, an effective equilibration between the iron core and the mantle after the Moon-forming event \citep{DeguenEtal2011}, {the late accretion of $\sim$10\% planetesimals} \citep{OlsonSharp2023}, or multiple collisions in a hit-and-run scenario \citep{AsphaugEtal2021}{, would reduce the concentration $^{182}$W in the mantle. Because Venus never experienced violent impact events such as the Moon-forming giant impact in our model, we expect the Venus mantle to be largely primordial, with a much higher $^{182}$W abundance than in Earth's mantle.}  

An early formation scenario, as proposed here, would resolve the challenges posed by the later formation model. These include: 1) {The absence of a moon for Venus is puzzling \citep{AsphaugEtal2021, MalamudPerets2024}, giving the prevalence of giant impacts in the late accretion model \citep{JacobsonEtal2014}}. 2) The need for post-impact damping mechanisms \citep{ClementEtal2023}, such as by leftover planetesimals, to reduce the eccentricities to their observed values \citep{Hansen2009}. Collisional growth also results in random orientation of the spin axis \citep{Chambers2001, MiguelBrunini2010}, inconsistent with solar terrestrial planets. {3) The magma ocean may exist and prevent the formation of an early and stable crust due to the frequent collisions in the late collisional growth scenario, inconsistent with the early crust on Earth \citep{Harrison2009}. }

Mars grew up in the first 1-10 Myr \citep{DauphasPourmand2011, KobayashiDauphas2013, MarchiEtal2020, WooEtal2021}, contemporaneous with the lifetime of the gaseous protoplanetary disk \citep{WeissEtal2021}. There are no fundamental reasons why growth for the other terrestrial planets would have been stalled at Mars size. Mature exoplanets are widely believed to be formed in the gas disk \citep{DrazkowskaEtal2023, MordasiniEtal2015}, as evidenced by the detections of young planets \citep{DavidEtal2019, KepplerEtal2019, BarberEtal2024, DaiEtal2024}.

Yet $>$80\% of those found by Transit (using Kepler and TESS) are not in resonance \citep{HuangOrmel2023i, HamerSchlaufman2024}. Various scenarios have been proposed to explain the overall observed non-resonant planetary architecture statistically, due to dynamical instability triggered by disk dispersal \citep{IzidoroEtal2021}, and planet-planet scattering \citep{WuEtal2024i, LiEtal2024}. 
Recently, radial velocity follow-up studies have revealed the existence of outer gas giant(s) to systems hosting inner planets with irregular architecture \citep{HeWeiss2023}. 
This finding aligns with the scenario that an external perturber destroyed the inner resonance architecture.  
{Scenario similar to the one proposed here for the Solar System could therefore have been common in the evolution of exoplanet systems \citep{YiEtal2025}. Notably, no outer giant planet has yet been observed in evolved resonance chain systems \citep[e.g., TRAPPIST-1][]{BossEtal2017}. }

\section{Conclusions}
\label{sec:conclusion}
We hypothesize that terrestrial planets (including Theia) in our Solar system formed early in the protosolar nebula. Naturally, they are trapped in a resonance chain. Starting with the resonant configuration, we use N-body simulations to study the destabilization of the resonance chain and long-term evolution. During the evolution, Theia commonly collided with proto-Earth (at a rate $20\sim50\%$, depending on the initial resonance between Earth and Theia). A comparison with the properties of Solar System terrestrial planets shows that our model satisfies four key constraints:
\begin{enumerate}
    \item The Mars-Venus period ratio of 3.05 is a relic of the former resonance chain. This value arises naturally in the early formation scenario, with a probability density an order of magnitude higher than in the late formation model (\fg{period_ratio_comp}). 
    \item The resulting planet eccentricities and mutual inclinations are moderate in the simulated systems. Their values are consistent with the current solar system (\fg{AMD_PMPV} upper panels and \fg{inc_ecc}). No post-moon forming damping is required. 
    \item The impact velocity between Earth and Theia is always ${\lesssim}1.5$ of the escape velocity (\fg{AMD_PMPV} lower panels) in our simulated systems, consistent with the Moon-forming giant impact \citep{TimpeEtal2023}. 
    \item The Earth-Theia giant impact follows the onset of giant planet instability, and occurs $\gtrsim10$ Myr afterwards (\fg{AMD_PMPV} lower panels). It is in line with cosmochemical dating. 
\end{enumerate}

{Two immediate predictions follow from our investigation. First, a giant impact event with Venus is disfavored in our model. Therefore, its mantle composition stays largely primordial, and we expect its isotopic signature to reflect early formation. Future missions characterizing isotopic abundances on Venus would offer the opportunity to verify the early formation models for the Solar terrestrial planets in this work. }

{Second, most exoplanets are not in resonance. They may start in resonance but later get destabilized. We demonstrate that the outer massive companion can destroy the resonance chain. Our dynamical study hints that those exoplanet systems that harbor evolved long resonance chains are unlikely to host external gas giants. 
}

\begin{acknowledgments}
We thank the anonymous referee for their constructive suggestions and comments. We would like to thank Anders Johansen, Beibei Liu, and Shoji Mori for their useful discussions. S.H. and C.W.O. are supported by the National Natural Science Foundation of China under grant Nos. 12250610189, 12233004, and
12473065. 

\end{acknowledgments}

\begin{contribution}

S.H. proposed this idea and conducted the numerical simulations. S.H. and C.W.O. initiated the collaboration. All authors contributed to analysing and discussing the numerical results and revising the manuscript. 


\end{contribution}

%

\software{Matplotlib \citep{CaswellEtal2021},  
          Rebound \citep{ReinLiu2012}, 
          Reboundx \citep{TamayoEtal2020}
          }


\appendix
\section{Formation of a resonant chain in the pebble accretion and planetesimal accretion models}
\label{ap:formation}

\begin{figure}
	\centering
	\includegraphics[width=0.48\linewidth]{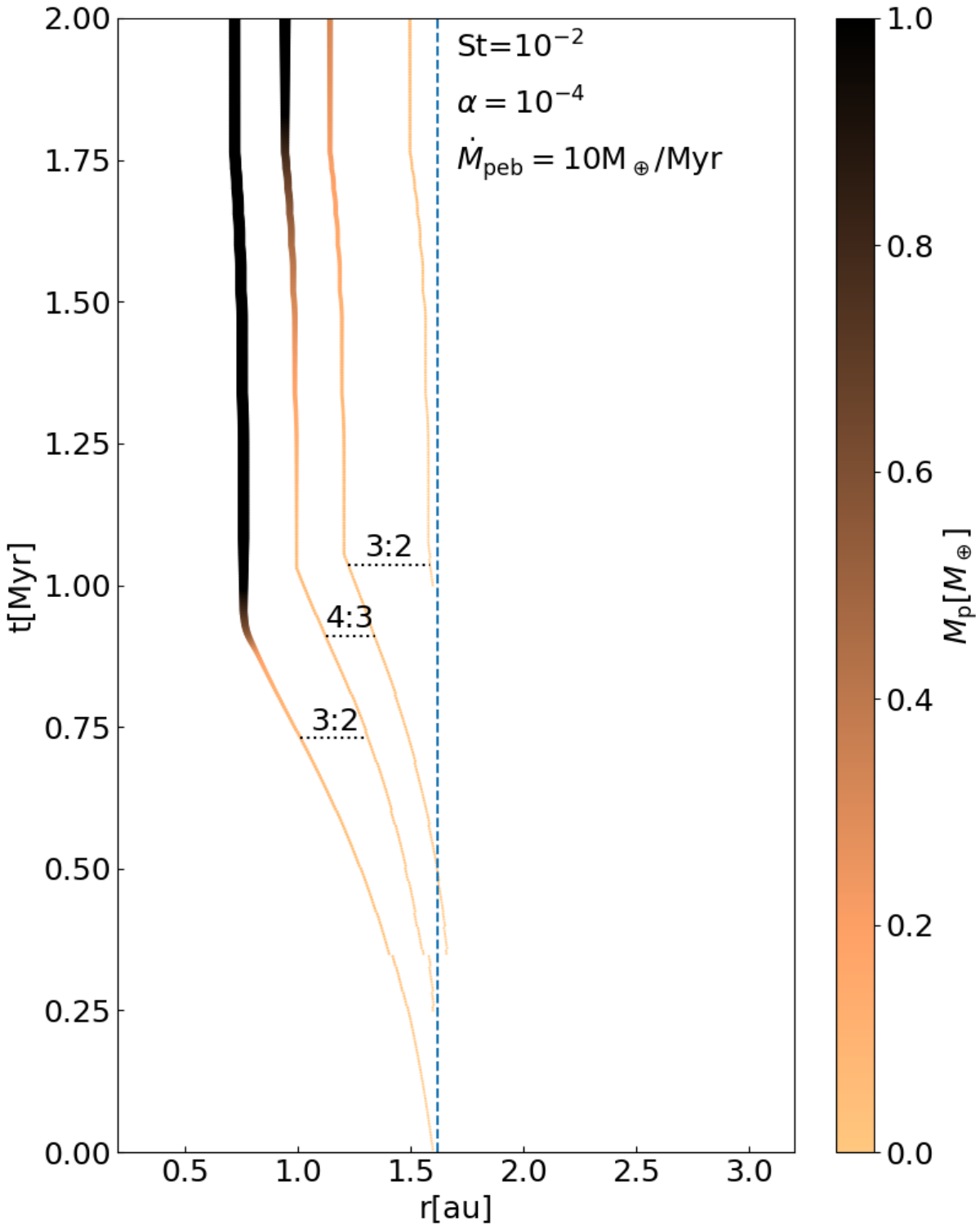}
	\includegraphics[width=0.48\linewidth]{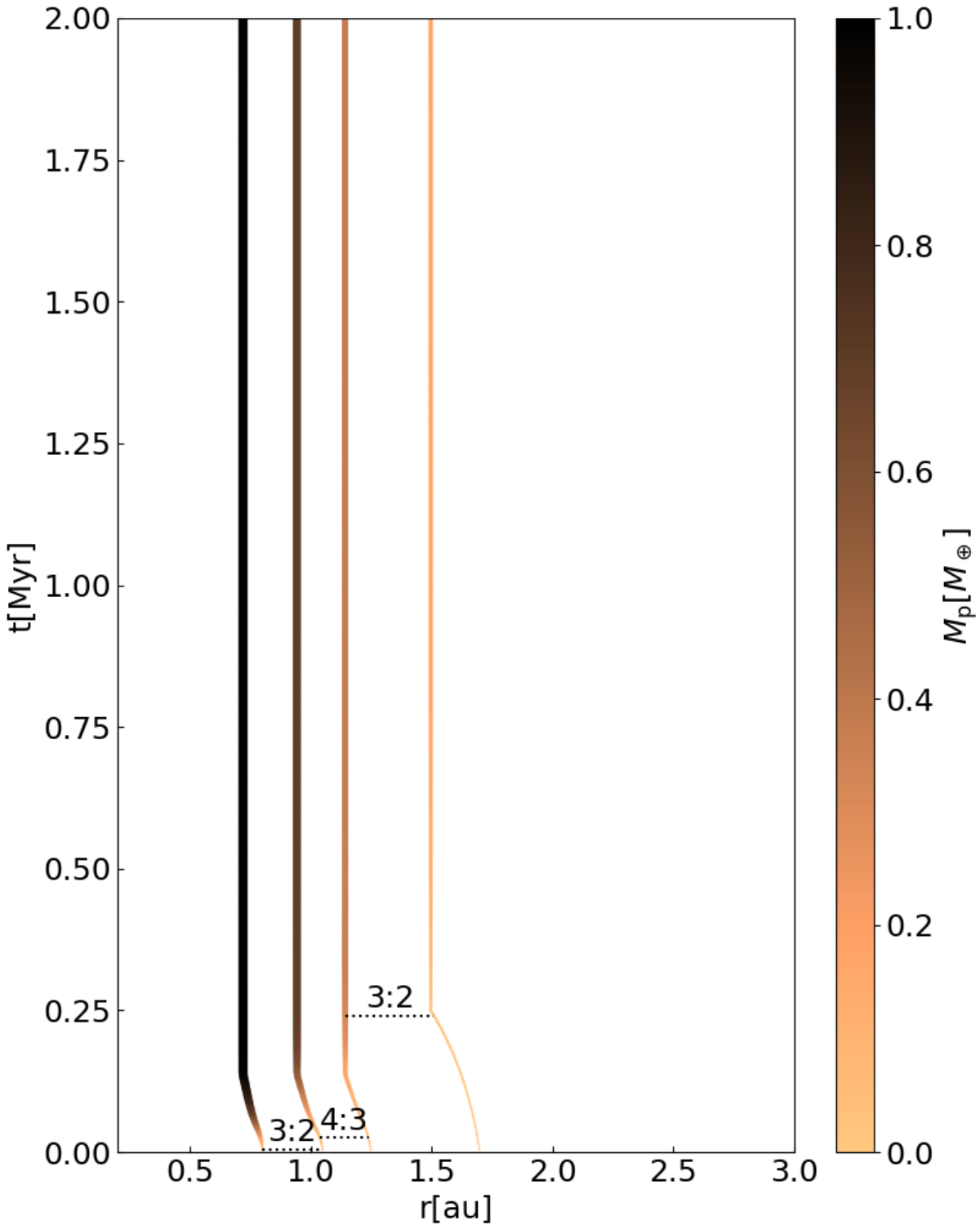}
	\caption{{Simulations of planetary accretion and migration, illustrating the formation of resonant architectures.} The left panel shows planets growing via pebble accretion, while the right panel depicts growth through planetesimal accretion, both occurring within a gaseous protoplanetary disk. In each panel, the trajectories represent the evolution of the semimajor axis (x-axis) and planetary mass (copper color) over time (y-axis). They denote, from left to right, Venus, Earth, Theia, and Mars. In the left panel, the vertical blue line marks the water snowline.  }
	\label{fig:planet_formation}
\end{figure}

We employ simplified planet formation models to demonstrate that the initial resonant architecture of the terrestrial planets is feasible. In \fg{planet_formation}, we show that both pebble accretion and planetesimal accretion can produce a resonant configuration within a gaseous disk—the starting condition for our study. In the pebble accretion scenario, planet embryos are introduced sequentially at the water iceline, then grow and migrate inward one after another. In contrast, the planetesimal accretion scenario initializes embryos simultaneously at different orbital distances. Ultimately, these processes yield a 2:3:4:6 resonance chain corresponding to Venus, Earth, Theia, and Mars. Notably, the final resonant architecture depends on the time interval (for pebble accretion) or spatial separation (for planetesimal accretion) between adjacent planet pairs. To achieve a tighter resonance between Theia and Earth, the interval between the initial Earth and Theia embryos must be reduced.

In the following, we detail the models and the parameter choices used for planet formation. We fix the gas disk surface density distribution at:
\begin{equation}
	\label{eq:gas}
	\Sigma_\mathrm{g}=\Sigma_\mathrm{g,0}f_\mathrm{g}\left(\frac{r}{1\,\mathrm{au}}\right)^{\beta_0}\left(1-\sqrt{\frac{r}{R_\mathrm{in}}}\right)e^{-\left(\frac{r}{R_\mathrm{out}}\right)^{2-\beta_0}},
\end{equation}
where $\Sigma_\mathrm{g,0}=2\,400\,\mathrm{g/cm^2}$ is the surface density in the minimum-mass solar nebulae (MNSN), $f_\mathrm{g}=5$ is the scaling factor of the initial gas surface density with respect to MNSN, $R_\mathrm{out}=30\,\mathrm{au}$ is the disk outer radius, and $\beta_0=-1.5$ is the surface density power law index. The disk inner radius is set to be $R_\mathrm{in}=0.6\,\mathrm{au}$, which prevents Venus from migrating further inward. Such an effective inner disk edge (migration barrier) could be triggered by the wind-driven accretion increasing with time \citep{OgiharaEtal2024i, WuChen2025}. It ensures the convergent migration of terrestrial planets \citep{BrovzEtal2021, ClementEtal2021}.
The disk temperature profile is fixed at
\begin{equation}
	T=T_0\left(\frac{r}{1\,\mathrm{au}}\right)^{\xi_0},
\end{equation}
we take the temperature power law index $\xi_0=-0.5$ and the temperature at 1 au $T_0=200\,\mathrm{K}$.

\begin{figure}
	\centering
	\includegraphics[width=0.5\linewidth]{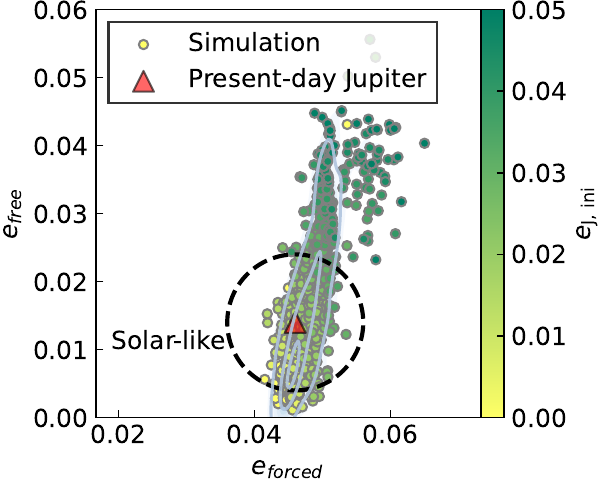}
	\caption{{Jupiter's free and forced eccentricities obtained at the end of the simulations.} The scatters with different colors are the results of the simulations starting with different initial Jupiter eccentricities. The solar system is marked as a triangle. The scatters in the circle (of radius 0.01) are indicated as the Solar-like systems in the simulation because they have similar Jupiter eccentricities as the present-day Jupiter.
	}
	\label{fig:fin_ej}
\end{figure}

For type-I migration, we use,
\begin{equation}
	\dot{r}=-k_\mathrm{mig}\frac{m}{m_\star}\frac{\Sigma_\mathrm{g}r^2}{m_\star}h^{-2}v_\mathrm{K},
\end{equation}
where $\dot{r}$ is the shrinking rate of the planet's semimajor axis, $h$ is the disk aspect ratio, and we take the mean molecular weight as 2.4. The prefactor $k_\mathrm{mig}$ depends on the disk properties. We adopt the fit from \citet{D'AngeloLubow2010},
\begin{equation}
	k_\mathrm{mig}=2(1.36-0.62\beta-0.43\xi),
\end{equation}
where $\beta$ and $\xi$ are the local gradient indices of disk surface density and temperature. As the two adjacent planets convergently migrate to the resonance location, they get trapped with their periods near integer commensurability.

If planets grow by pebble accretion (\fg{planet_formation} left panel), we can get the accretion rates of the planet via
\begin{equation}
	\dot{m}=\epsilon \dot{M}_\mathrm{solid},
\end{equation}
The 3D pebble accretion efficiency $\epsilon$ is calculated following \citet{OrmelLiu2018}. It depends on the planet’s eccentricity, the particle Stokes number St, the local gas pressure gradient $\eta$, the disk aspect ratio $h$, and the gas turbulent diffusion parameter $\alpha_\mathrm{t}$. The quantities $h$ and $\eta$ follow from the disk structure. We assume zero eccentricities, fix St=0.01 and $\alpha_\mathrm{t}=10^{-4}$, and assume an incoming pebble mass flux of $\dot{M}_\mathrm{dust}=10 \, m_\oplus\,\mathrm{Myr}^{-1}$.

{One of our simulation results is shown in \fg{planet_formation} left panel. Similar to \cite{JohansenEtal2021}, we initialize four planet embryos at the iceline sequentially. They grow and migrate inward one by one. Venus stops its migration at 1 Myr because it reaches the migration barrier \citep{OgiharaEtal2024i, WuChen2025}. As the latter planets join, they form a 2:3:4:6 resonance chain. Their mass also increases after the formation of a resonance chain due to pebble drift in the disk. 
At the end of the simulation time (2 Myr), all four planets have their mass close to the present values. }

\begin{figure}
	\centering
	\includegraphics[width=\linewidth]{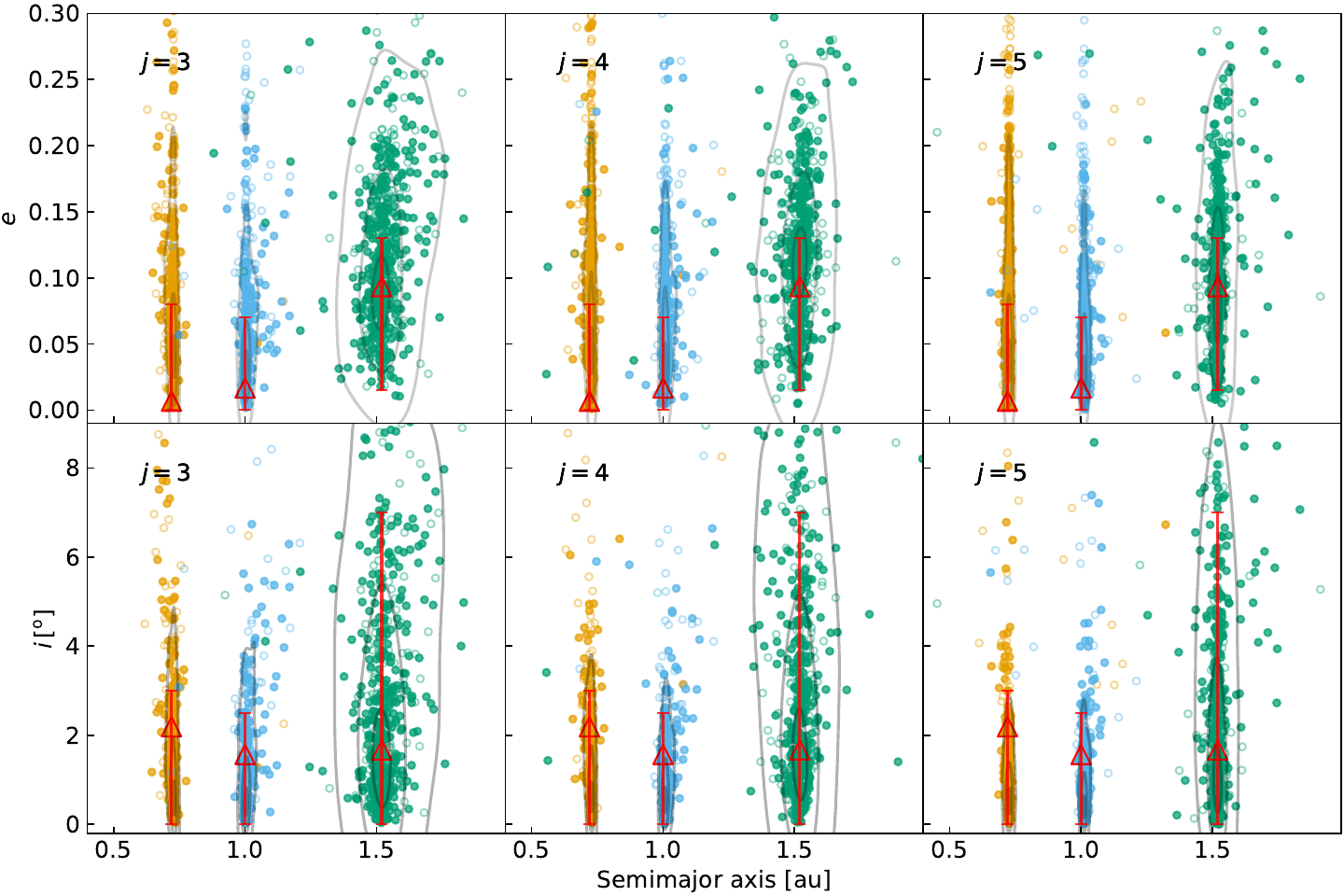}
	\caption{Dynamical properties for planet systems, in which Theia collides with Earth. The upper panels show the final eccentricities of Venus, Earth, and Mars, while the lower panels display their final inclinations relative to the invariable plane. Different scatter points are the results of the simulations that end up with Solar-like (solid dots) or other (open dots) systems (see \fg{fin_ej} for definition). We overplot the median 10\%, 50\%, and 90\% contours of the scatter. Each column corresponds to a different initial Venus-Mars resonance. For reference, we include the present-day eccentricities of the terrestrial planets and the inclinations (triangle) and the range of them on secular timescales (error bar) of the terrestrial planets \citep{ItoTanikawa2002}. }
	\label{fig:inc_ecc}
\end{figure}

Alternatively, if planets grow up from accreting planetesimals (\fg{planet_formation} right panel), the planet oligarchic growth timescales follow from \citep{KokuboIda2002}
\begin{equation}
	\tau_\mathrm{p} = 1.2\times10^5 \, \mathrm{yr}\left(\frac{\Sigma_\mathrm{plts}}{10\,\mathrm{g}\,\mathrm{cm}^{-2}}\right)^{-1}\left(\frac{r}{1 \, \mathrm{au}}\right)^{\frac{3}{5}}\left(\frac{m}{m_\oplus}\right)^{\frac{1}{3}} \left(\frac{\Sigma_\mathrm{g}}{2400\, \mathrm{g}\,\mathrm{cm}^{-2}}\right)^{-\frac{2}{5}}\left(\frac{m_\mathrm{plts}}{10^{18}\, \mathrm{g}}\right)^{\frac{2}{15}},
\end{equation}
where $m$ is planet mass, the mass of a planetesimal $m_\mathrm{plts}$ is fixed to be $10^{18}$ g, equivalent to a diameter of $\approx$6 km for an internal density of $1$ g/cm$^3$. The initial planetesimal disk surface density distribution $\Sigma_\mathrm{plts}$ follows a similar shape as the gas disk in \eq{gas},
\begin{equation}
	\Sigma_\mathrm{plts}=90\,\mathrm{g/cm^2}\left(\frac{r}{1\,\mathrm{au}}\right)^{\beta_\mathrm{plts}}\left(1-\sqrt{\frac{r}{R_\mathrm{in}}}\right)\exp\left[-\left(\frac{r}{R_\mathrm{out}}\right)^{2-\beta_\mathrm{plts}}\right],
\end{equation}
but with most of the planetesimal concentrated at $\sim1$ au by using a much steeper gradient value $\beta_\mathrm{plts}=-5.5$. Such a ring-like distribution of planetesimals is advantageous for forming a small Mars, aligning with the previous findings \citep{WooEtal2024}. Similar to  \citet{EmsenhuberEtal2021}, we subtract the accreted planetesimal mass from the planetesimal surface density profile. The width of the feeding zone is 6 times the planet Hill radius centered at the planet orbit \citep{Lissauer1993}.

{The planet formation simulation via the above planetesimal accretion is shown on the right panel of \fg{planet_formation}. All four embryos are initialized at the start, but at different locations. The ring-like distribution of planetesimals is centered at 1 au, promoting the growth of Venus and Earth. Therefore, they gain more mass than Theia and Mars. Finally, they consume all the planetesimals in their feeding zones, reaching the planetesimal isolation mass. The planets are born inside the 2:1 resonance. With slight migration, they are trapped in the 2:3:4:6 resonance chain. }

Despite simplifying processes such as disk evolution and planetesimal scattering, and lacking exhaustive parameter exploration, these simulations reveal that both planetesimal and pebble accretion models yield resonant configurations comparable to the initial conditions of our post-disk N-body framework. 

\section{Parameter sensitivity}
\label{ap:parameter}

\begin{figure}
	\centering
	\includegraphics[width=0.8\linewidth]{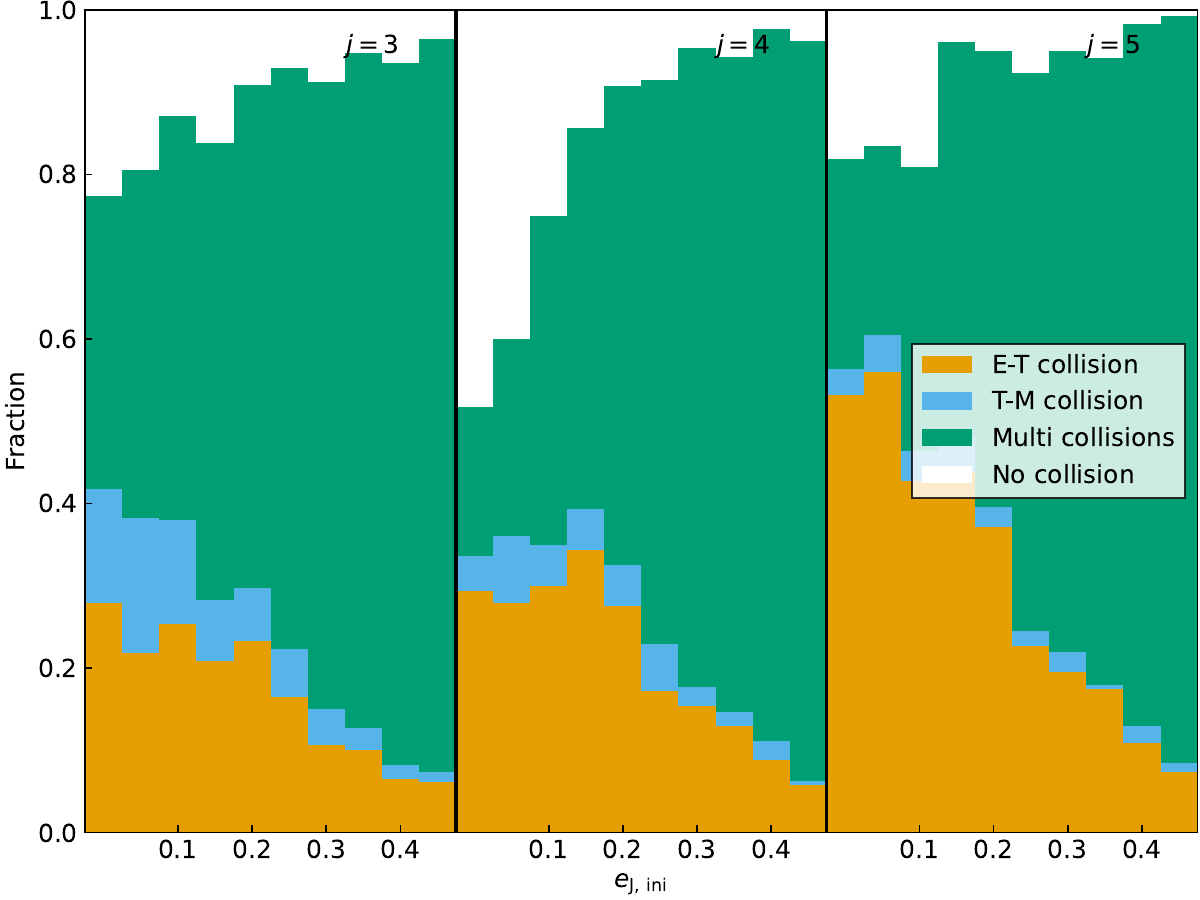}
	\caption{{Statistical outcomes of the simulated Solar Systems and their dependence on initial conditions.} The architecture of simulated systems is classified into four groups: 1) Theia collides with Earth (orange); 2) Theia collides with Mars (blue); 3) There are multiple collisions (green); 4) No collisions occurred within the end time of the simulation (white). The left, middle, and right panels display results for different initial resonances between Earth and Theia (the resonance number is shown at the top right of each panel). Within each panel, separate histograms represent simulation outcomes starting with different initial eccentricities for Jupiter. 
    }
	\label{fig:full_success}
\end{figure}

To assess how the properties of the simulated systems depend on the initial conditions, we systematically varied two key parameters: the initial eccentricity of Jupiter $e_\mathrm{J,\,ini}$, and the initial resonance configuration between Earth and Theia, characterized by the resonance number $j$. In contrast to these parameters, our experiments show that the results are nearly insensitive to variations in the eccentricity damping parameter in the disk $K_\mathrm{e}$ and Saturn’s outward migration timescale $\tau_\mathrm{S}$. The timescale of Saturn's outward migration $\tau_\mathrm{S}$ is parameterized between 0.8 to 16 Myr. The eccentricity damping factor $K_\mathrm{e}$ is changed logarithmically in a range of evenly-spaced grids from $10^2$ to $10^4$.

Naturally, varying the initial eccentricity of Jupiter affects its final eccentricity. In \fg{fin_ej}, we show the forced and free eccentricities from all our simulations. We change $e_\mathrm{J,\,ini}$ in the range of 0 to 0.05 (linear grid). As $e_\mathrm{J,\,ini}$ increases, both $e_\mathrm{forced}$ and $e_\mathrm{free}$ rise, with $e_\mathrm{free}$ exhibiting a more pronounced increase. For comparison, we also plot the observed value of Jupiter’s eccentricity. Our results indicate that simulated systems align best with the Solar System when $e_\mathrm{J,\,ini}<0.02$. 

{We select the simulated systems with their final Jupiter eccentricity close to the present-day Jupiter. Those systems are labeled as "Solar-like systems", as highlighted by the circle in \fg{fin_ej}. The radius of the circle is 0.01. Then, the dynamical properties of the planets in the Solar-like systems are compared with those in the real Solar system. We already studied the angular momentum deficit (AMD), period ratio, impact velocity, and the timing between Theia and Earth in the main text (\fg{AMD_PMPV}). Here, the eccentricity and inclination of each terrestrial planet are shown in \fg{inc_ecc}. The eccentricity and inclination values in our simulated systems are consistent with the secular evolution of the real Solar system. To give a full comparison, the information of those systems that are not Solar-like is also exhibited using open circles in \fg{inc_ecc}. } We also test three different initial architectures, corresponding to three different values of the resonance number $j=3,4,5$ between Theia and Earth. As $j$ increases, Theia stays in an orbit closer to the Earth. As a result, Mars experiences less close encounters with Theia and its semimajor axis is less dispersed.

Variations in the initial eccentricity of Jupiter, $e_\mathrm{J,\,ini}$, lead to markedly different system architectures (see \fg{full_success}). A higher $e_\mathrm{J,\,ini}$ more strongly perturbs the inner terrestrial planets via secular interaction, resulting in a higher frequency of multiple collision events. In contrast, when Jupiter starts on a nearly circular orbit, such violent interactions are less common. A higher fraction of systems only have one single collision or are even collision-free over the 100 Myr simulation period. In the systems that stayed collision-free, half have their resonance angles circulate. For those systems, we expect the planets to eventually collide. Otherwise, the terrestrial planets are still linked by the original resonance chain. As $j$ increases, more and more simulated systems have Theia and Earth collisions (orange bars) while fewer have Theia colliding with Mars (blue bars). In \fg{full_success}, the number of collision-free systems increases when $j$ increases from 3 to 4, due to the stronger resilience of the 5:4 resonance. As $j$ increases to 5, the number of collision-free systems decreases again due to the instability triggered by too close orbits between Theia and Earth \citep{PetitEtal2020i}.

\begin{figure}
	\centering
	\includegraphics[width=0.8\linewidth]{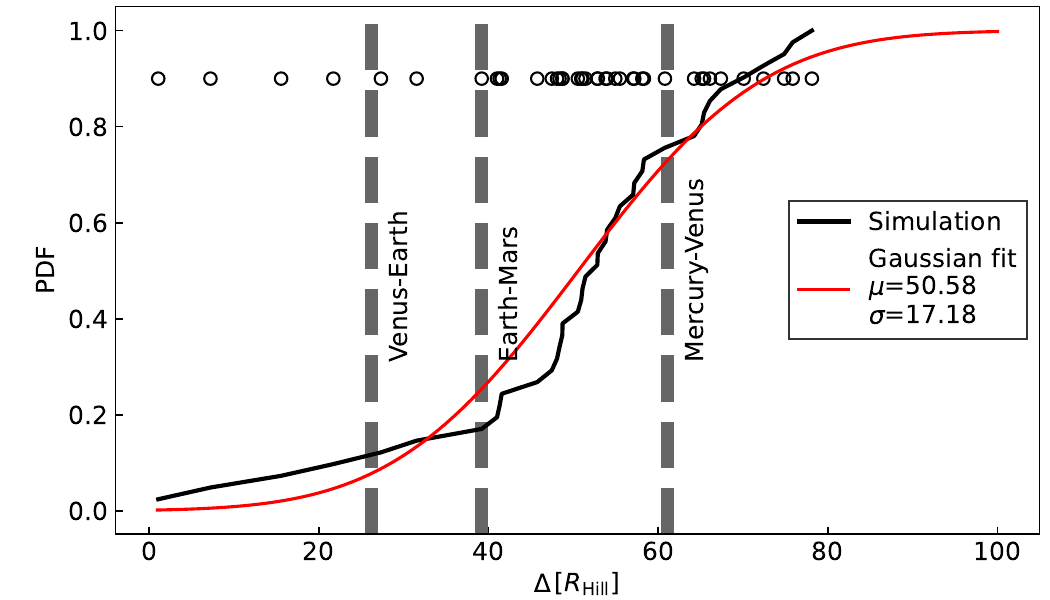}
	\caption{{Cumulative distribution of the mutual separation ($\Delta$) between each two adjacent terrestrial planets in the collisional growth scenario.} The unit is the mutual Hill radius. The black curve shows the distribution of the simulation results. Each data point from the simulation is shown with the black circles. We use a Gaussian to fit the probability distribution, as shown in the red curve. The vertical dashed lines show the values from the terrestrial planets in the Solar system. }
	\label{fig:sim_fit}
\end{figure}

\begin{figure}
	\centering
	\includegraphics[width=0.8\linewidth]{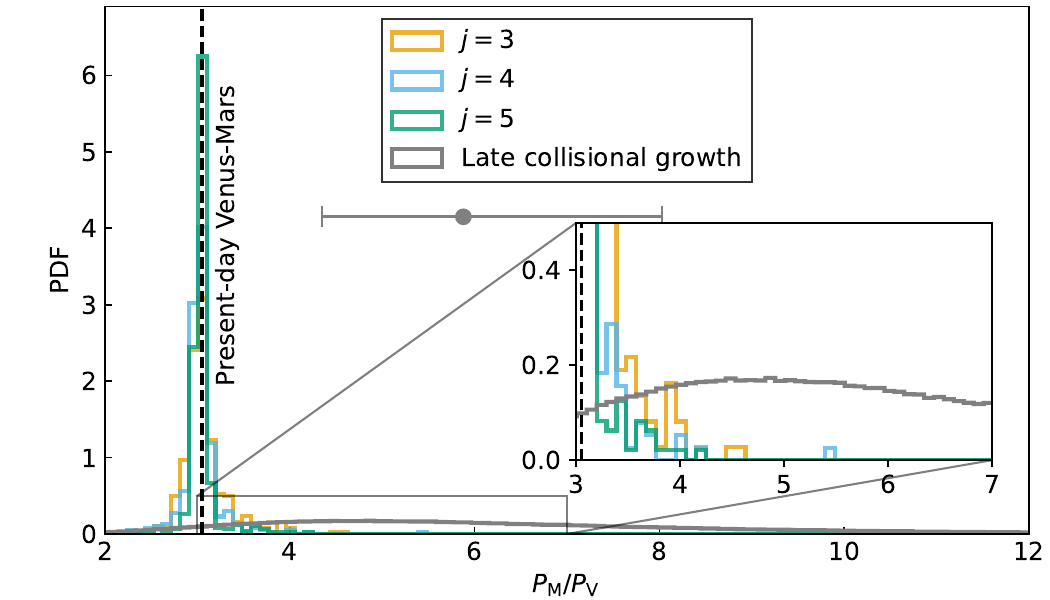}
	\caption{{Mars-Venus period ratio probability distribution from different models}. Similar to \fg{period_ratio_comp}, here we zoom in to display the shape of the period ratio distribution resulting from the late formation scenario (grey histogram).{ The median 50\% range and the median value of the late formation distribution are indicated by the error bar and its center value.} We zoom in on the plot in the inset such that the peak of the period ratio distribution in the late formation scenario is visualized.  }
	\label{fig:period_ratio_comp_zoom}
\end{figure}

\section{Venus-Mars period ratio in the late collisional growth scenario}
\label{ap:latecollision}
While exhibiting the properties of simulated systems in our early formation scenario, we also compare the Venus-Mars period ratio with that from the late formation scenario. In such a case, planets are formed by the collisional growth between planetesimals and Mars-sized embryos. 

The simulation setups for the late collisional growth scenario are the same as those used in \citet{KokuboPrep}. We give a brief description here. The simulations start with 15 protoplanets of equal mass, each with $m=0.15\,m_\oplus$. Their semimajor axes are distributed between 0.5 and 1.5 au, with adjacent protoplanets separated by 10 times the mutual Hill radius. The initial eccentricities and inclinations follow Rayleigh distributions with dispersions of $\sigma_e=0.01$ and $\sigma_I=0.005$. Each body is assumed to have a mean bulk density of $3\,\mathrm{g\,cm^{-3}}$. In total, 20 independent simulations are performed.

The dynamical evolution of the system is computed using a modified Hermite integrator, as described by \citet{KokuboPrep}, which efficiently resolves close encounters while ensuring accurate energy conservation in gravitational N-body simulations. To accelerate the simulation, a block timestep scheme has been adopted \citep{Makino1991}. Perfect accretion is assumed when two planets collide, conserving both mass and momentum. Each simulation runs for 1.7$\times10^{2}$ Myr.

We here calculate the separation between each pair of adjacent planets at the end of the 20 simulations, and the results are shown in \fg{sim_fit}. This figure shows the distribution of the post-giant impact semi-major axes difference among the planets in units of the mutual Hill radius ($\Delta$). Systems with close-to-zero $\Delta$ are allowed (stable) due to their large mutual inclinations, $\sim10^\mathrm{o}$. The separations among the terrestrial planets in the Solar system are also indicated. The simulated distribution of separations closely matches the values observed in the Solar system. We find that the probability distribution peaks around 50 Hill radii. To fit this distribution, we use a Gaussian profile. The best fit to the distribution in $\Delta$ is given by a mean value of $\mu = 50.58$ and a standard deviation $\sigma = 17.18$.

Using this separation distribution, we apply the Monte Carlo method to generate a distribution of the period ratio between Venus and Mars. We generate two sets of separation distances: one covering the Venus-Earth pair and the other for the Earth-Mars pair, assuming that $\Delta\sim\mathcal{N}(\mu,\sigma)$ for each of them. The set of synthetic $\Delta$ is then converted back into period ratios, and the resulting distribution is shown in \fg{period_ratio_comp}. \Fg{period_ratio_comp_zoom} provides the same information, but with an extended period ratio. The peak of the period ratio distribution of the late formation scenario ($\approx4.9$) is about twice that of the present-day value (3.05). Statistically, only the lower 10\% percentile covers the real Mars-Venus period ratio of 3.05. More detailed analysis will be presented in \citep{KokuboPrep} in the context of dynamical stability criteria for Kepler planet systems.

\bibliography{ads}{}
\bibliographystyle{aasjournal}


\end{CJK*}
\end{document}